\documentclass{SciPost}
\pdfoutput=1
\usepackage{graphicx}
\usepackage{mathtools}

\renewcommand{\v}[1]{{\bf {#1}}}%
\newcommand{\kk}{\v{k}}%
\newcommand{\figla}[1]{\v{#1}}%
\newcommand{\qq}{\v{q}}%
\newcommand{\KK}{\v{K}}%
\newcommand{\RR}{\v{R}}%
\newcommand{\rr}{\v{r}}%
\newcommand{\bl}{\v{l}}
\newcommand{\QQ}{\v{Q}}%
\newcommand{\uu}{\v{u}}%
\newcommand{\nn}{{\rr}}%
\newcommand{\ttau}{\boldsymbol{\tau}}%
\newcommand{\ov}[1]{\vec{\boldsymbol{#1}}}
\newcommand{\mat}[1]{{\v{#1}}}
\newcommand{\HH}{{\mathcal{H}}}%
\renewcommand{\vec}[1]{{{#1}}}
\newcommand{\pdag}{{\phantom{\dag}}}%

\begin{document}

\begin{center}{\Large \textbf{
Creating better superconductors by periodic nanopatterning
}}\end{center}

\begin{center}
M.P. Allan\textsuperscript{1*}, 
M.H. Fischer\textsuperscript{2,3}, 
O. Ostojic\textsuperscript{1},
A. Andringa\textsuperscript{1}
\end{center}

\begin{center}
{\bf 1} Leiden Institute of Physics, Leiden University, Niels Bohrweg 2, 2333 CA Leiden, The Netherlands
\\
{\bf 2} Department of Condensed Matter Physics, Weizmann Institute of Science, Rehovot 7610001, Israel
\\
{\bf 3} Institute for Theoretical Physics, ETH Zurich, 8093 Zurich, Switzerland

* allan@physics.leidenuniv.nl
\end{center}

\begin{center}
\end{center}


\section*{Abstract}
{\bf 
 The quest to create superconductors with higher transition temperatures is as old as superconductivity itself.
One strategy, popular after the realization that (conventional) superconductivity is mediated by phonons, is to chemically combine different elements within the crystalline unit cell to maximize the electron-phonon coupling. This led to the discovery of NbTi and Nb$_3$Sn, to name just the most technologically relevant examples.
Here, we propose a radically different approach to transform a `pristine' material into a  better (meta-) superconductor by making use of modern fabrication techniques: designing and engineering the  electronic properties of thin films via periodic patterning on the nanoscale.  
We present a model calculation to explore the key effects of different supercells that could be fabricated using nanofabrication or deliberate lattice mismatch, and demonstrate that specific pattern will enhance the coupling and the transition temperature. 
We also discuss how numerical methods could predict the correct design parameters to improve superconductivity in materials including Al, NbTi, and MgB$_2$
}

\section*{}
Conventional --- i.e., phonon-mediated --- superconductors include many elemental metals with transition temperatures between {1}{K} and {10}{K}, simple alloys like NbTi and Nb$_3$Sn with transition temperatures up to {$\sim${20}{K}}, and MgB$_2$ with a record transition temperature of {39}{K} at ambient pressure~\cite{Buzea2001}. Mainly because  of superior material properties that make fabrication and handling easy, these materials have widespread technological applications, ranging from medical magnetic resonance imaging to quantum information technologies.  The effort  to improve the  quality of these conventional superconductors for applications has all but stopped with  the discovery of high-temperature superconductors~\cite{Rogalla2011,Marsiglio2008,Ginzburg1968} (with notable exceptions~\cite{Pickett2006,Pickett2008,Karakonstantakis2013}).   Yet, any improvement of the quality of conventional superconductors has immediate and wide-ranging technological impact.  

\begin{figure}[tb] 
\centering
      {\includegraphics[width=.9\textwidth]{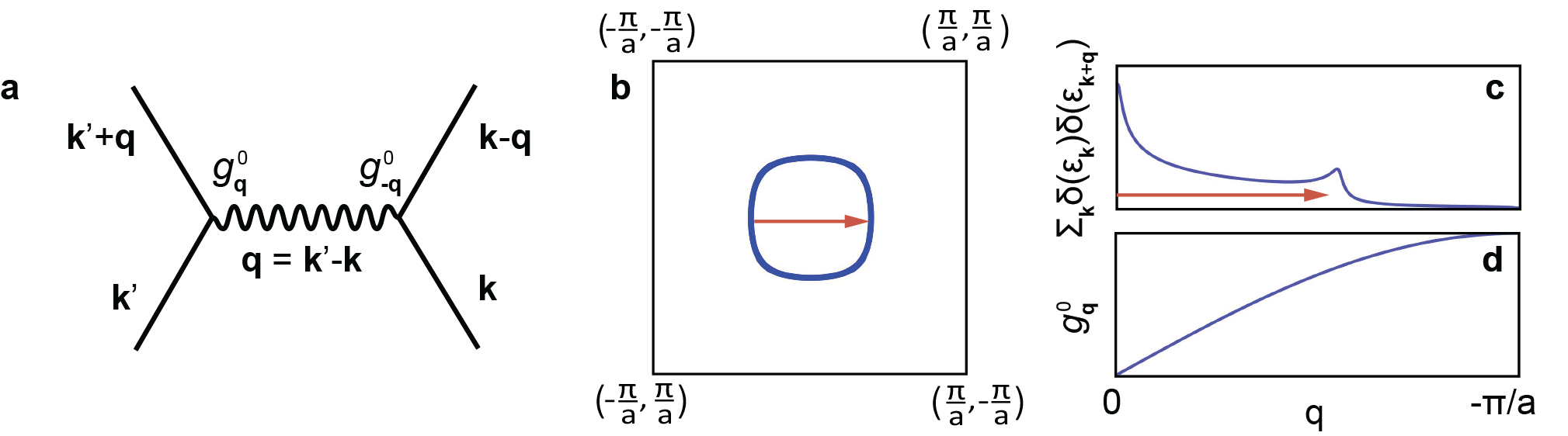}}
      \caption{Electron-phonon interaction and the electron-phonon coupling parameter $\lambda$. \figla{a},  Diagrammatic representation of the coupling between electrons with momentum $\kk$, $\kk'$  through the exchange of a  phonon with momentum $\qq$ and interaction matrix element $g_{\kk\qq}^0$ (here dependent only on $\qq$, $g^0_{\kk\qq}=g^0_{\qq}$). \figla{b},   Due to the kinematic constraint (Eq.~(\ref{eq:coupling})), only scattering vectors connecting the Fermi surface points are relevant (red arrow). \figla{c}, Kinematic constraint along a high-symmetry direction. \figla{d}, The structure of the interaction matrix element  determines what kinematic constraints lead to a  high electron-phonon interaction $\lambda$; in the example here a large Fermi surface is beneficial. }\label{fig:intro} 
\end{figure}

Here, we present a new method for such improvement using modern nanofabrication that allows for the creation of new materials with specially designed electronic (and phononic) structures that maximize the pairing interaction (Fig.~\ref{fig:intro}). In short, the idea is to engineer the phononic and electronic structure of a given material by introducing specific nanofabricated periodic supercell structures on thin films. In many ways, the strategy we propose here is  similar to creating new superconductors by chemically changing their unit cell as it has been done mainly in the middle of the 20th century~\cite{Marsiglio2008}, but approaching instead from the long-range limit, as current nanofabrication already allows structures of roughly 5 to 50 unit cells~\cite{Grigorescu2009, Genet2007,Schneider2010, Corso2004, Allan2007, Hunt2013, Eigler1990, Kalff2016}.

\begin{figure}[tb] 
\centering
      \includegraphics[width=.8\textwidth]{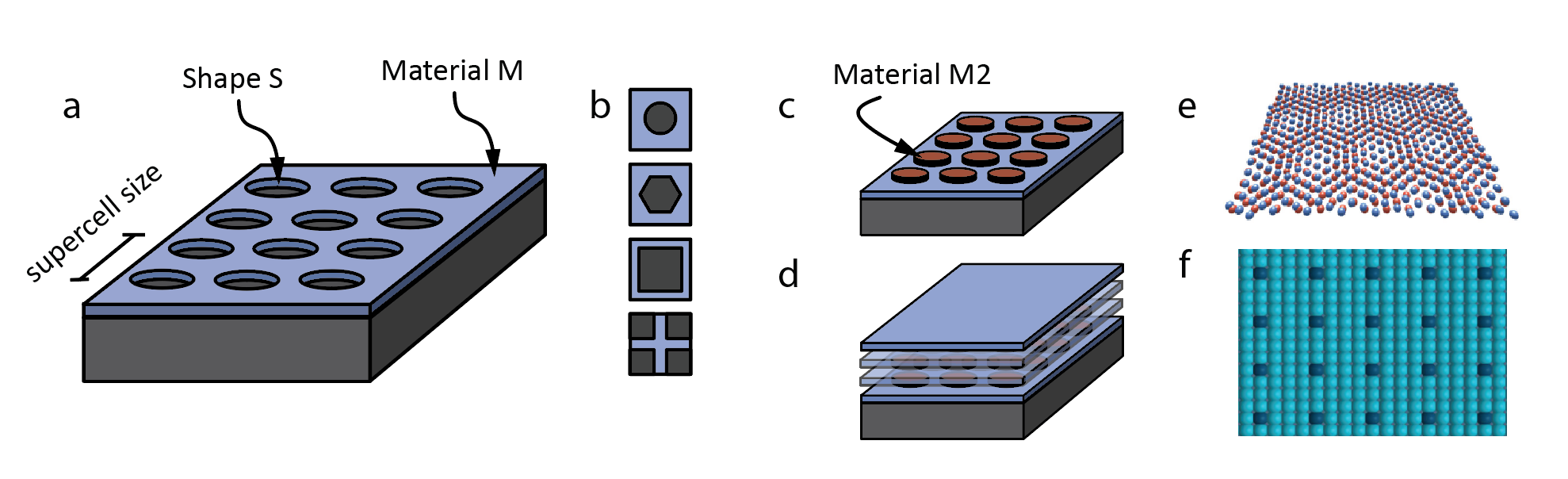}
      \caption{Fabrication methods. \figla{a}, Modern nanofabrication tools allow one to make nano-patterned shapes with supercell periodicities of $\lesssim$5 to 50 lattice constants. \figla{b}, Different shapes have different effects on the resulting electron-phonon coupling parameter $\lambda$. \figla{c}, Different layers of (insulating) materials on top of the thin films will influence the phonon and electron dispersions individually and can increase the phonon energies.  \figla{d}, Stacking allows for 3D materials.  \figla{e,f}, Smaller patterning are possible using Moir\'e engineering or single atom manipulation.}\label{fig:fab} 
\end{figure}

Before elaborating on the model calculation, we sketch methods to fabricate the  nanofabricated patterns (Fig.~\ref{fig:fab}). 
The starting point is a `pristine' material which is generally a thin film of a known superconductor (or any other material).   One can  pattern  the film  using standard cleanroom tools such as electron beam lithography~\cite{Grigorescu2009}, photolithography, or focused ion beam lithography with He or Gd ions~\cite{Genet2007}, to make a supercell of a given size and shape (Fig.~\ref{fig:fab}{b}-{d}). Using these methods it is possible to make periodic supercell structures down to a few nanometers in size, and even smaller patterns are possible using Moir\'e engineering (Fig. 2e)~\cite{Corso2004, Allan2007, Hunt2013}, self-assembly~\cite{Nie2010}, or atomic scale manipulation with scanning probe microscopy (Fig. 2f)~\cite{Eigler1990,Kalff2016}.  The choice of material, periodicity and supercell shape will allow for considerable freedom to design the desired electronic and phononic structure, together or individually.

The transition temperature of a conventional superconductor depends on the electronic and phononic structure as well as the coupling matrix elements in between them; the effect of all three is conveniently summarized in the dimensionless electron-phonon coupling parameter $\lambda$. In Bardeen-Cooper-Schrieffer theory~\cite{Bardeen1957,Parks1969}, the critical temperature depends exponentially on $\lambda$, $T_c\propto  \omega_D e^{-1/\lambda}$, where $\omega_D$ is the Debye frequency. $\lambda$ thus represent an ideal figure of merit. We can calculate $\lambda$ of the pristine material from the  interaction matrix element $g_{\kk\qq}^0$ for the scattering  of an electron with momentum $\kk$ to momentum $\kk+\qq$ and the electron  ($\varepsilon_\kk$) and phonon dispersions ($\omega_\qq$), by integrating over all possible scattering processes shown in Fig.~\ref{fig:intro}\figla{a}~\cite{Marsiglio2008,Parks1969, Allen1972,Yin2006}. In the adiabatic limit it is given by 
\begin{equation} 
    \lambda^{\rm{pristine}} =  \sum_{\kk,\qq}\frac{2}{\omega_{\qq} N(0)}  |g_{\kk\qq}^0|^2 \delta(\varepsilon_{\kk})\delta(\varepsilon_{\kk + \qq}).
    \label{eq:coupling}
\end{equation}
 The delta functions ensure that  only states at the Fermi level participate. The main ingredients for $\lambda$ are thus \emph{(i)}  the interaction matrix element of each process,  \emph{(ii)}  the electronic density of states at the Fermi level $N(0) \propto \sum_{\kk \qq}\delta(\varepsilon_{\kk})$, which determines the number of allowed processes, and  \emph{(iii)} the kinematic constraints from the Fermi surface and the coupling matrix element, given by the two delta functions and their matching with the momentum-space structure of the interaction matrix element (Fig.~\ref{fig:intro}\figla{b},\figla{c}).  (Note that all sums are understood to be conventionally normalized by the number of lattice sites.)
In the following, we discuss how to exploit these ingredients. 

We use a model calculation that contains the key knobs of the method to explore the opportunities of designed electron  dispersions for enhancing $T_c$ of a  pristine material with a given coupling strength $g_{\kk\qq}^0$, and to demonstrate the feasibility of the concept.
Our starting point is a two-dimensional pristine material defined on a square lattice. The electrons and   phonons are coupled through the  local shift of the  chemical potential an electron feels due to the deformation of the positively charged lattice background from a phonon. The resulting  interaction Hamiltonian reads~\cite{Coleman2015}
\begin{equation*}
    \HH_{\rm{int}} = {D} \sum_\nn (\vec{\nabla}\!\cdot\! \uu_\nn^\pdag) c^\dagger_\nn c_\nn^\pdag,   \\
\end{equation*} 
where $c_{\nn}^{\dag}$ creates an electron on lattice site $\nn$ and $\uu_\nn$ is the phonon displacement field. The proportionality $D$ indicates the  change of the chemical potential per volume change and is commonly called displacement potential. Note that in general ${D}$ is not a constant, but reflects the shape of the atomic potential. 
We further describe the electrons by a nearest-neighbor tight-binding model and use harmonic potentials for the phonons.   The details of the model are described in the Appendix. 

\clearpage
\begin{figure}[!tb] 
\centering
      {\includegraphics[width=1.\textwidth]{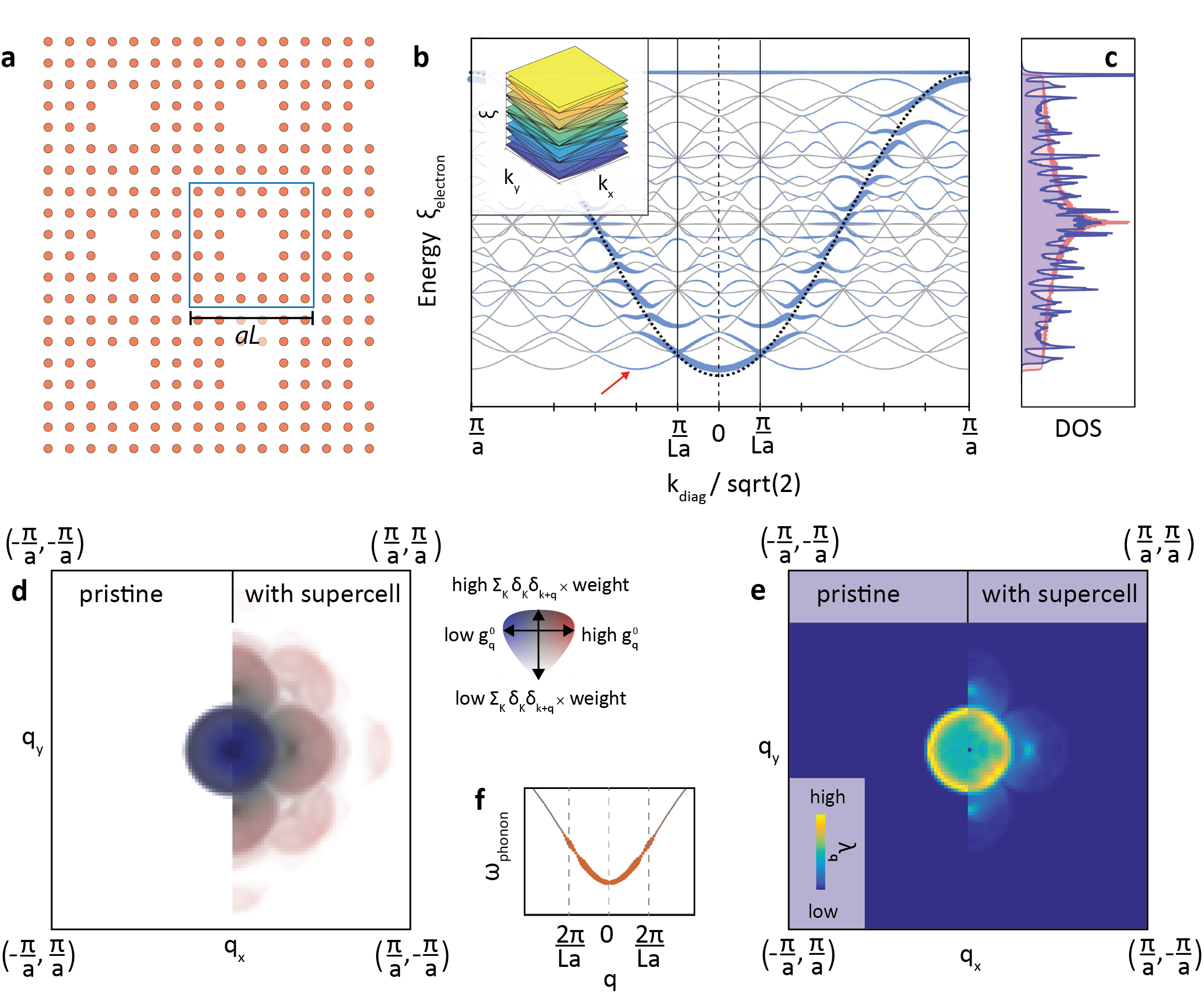}}
      \caption{Increased electron-phonon coupling parameter $\lambda$. \figla{a}, Illustration of our model  with a square lattice  and  a $6\times6$ supercell with a $2\times2$ hole; results for different supercell sizes and shapes are in Fig.~\ref{fig:con}. \figla{b}, The supercell periodicity reduces the size of the Brillouin zone by  a factor $L$ (black line). The extended Brillouin zone is plotted  as we  consider umklapp scattering between different zones. The bands $\xi^\nu_\kk$  are shown in gray, the blue color marks the `weight'. \figla{c}, Density of states of the pristine material (red) and the material with supercell (blue). \figla{d}, The phase space for scattering. The darkness indicates how much phase space is available, the red-blue indicates the strength of the interaction matrix element $g_{\kk\qq}^0=g_{\qq}^0$.   In the pristine material, only small scattering vectors are allowed due to the kinematic constraints; in the modified material, larger $\qq$, where the interaction is stronger (red) are possible.  \figla{e, f}, $\lambda_\qq$ as an indication of how much different modes couple, as a colorplot (\figla{e}) and along a high-symmetry direction (\figla{f}) where the width of the  line  indicates the contribution of a phonon with that wave vector.}\label{fig:model} 
\end{figure}

Next, we take the effect of the nano-patterned supercell into account and investigate how this allows to increase the coupling.  We concentrate on the electronic structure,  as in Bardeen-Cooper-Schrieffer theory, the details of the phonon dispersion has little effect on the critical temperature.   We use a supercell of $L \times L$ lattice sites and a hole in the center (Fig.~\ref{fig:model}\figla{a}) as a model of the realization shown in Fig.~\ref{fig:fab}\figla{a}, both because it is theoretically accessible and because it appears most promising.  Note, however, that within the model, we can include different forms of supercells.  For the electrons, the new supercell periodicity is  reflected in a reduction of the Brillouin zone  area by a factor of $L^2$ (Fig.~\ref{fig:model}\figla{b}) and $L^2$ back-folded bands ($\xi_\KK^\nu$) with band gaps at the reduced Brillouin zone  boundary ($\nu$ is the band index). Fig.~\ref{fig:model}\figla{b} shows the resulting bands for a $6\times 6$ supercell. 
These back-folded bands (``shadow bands'') can thus in principle help to overcome the kinematic constraints discussed above or in Eq.~(\ref{eq:coupling}). The scattering between the backfolded bands corresponds to umklapp scattering between different reduced Brioullin zones. The strength of these umklapp processes  that connect different states is determined by a `weighting' related to the transformation between old and new basis, which is closely related to the overlap of the new states with the states of the pristine material (denoted by blue lines in \ref{fig:model}\figla{b}). It is this weight that ensures that in the limit of large supercells, the enhancement of $\lambda$ must vanish. Choosing the shape and size of the supercell allows to  influence the weight and have the new scattering vectors align with the interaction matrix element of the pristine material at specific $\qq$ points.  Here, we  absorb the weighting into a new interaction matrix element $g^{\nu\nu'}_{\KK\qq}$. The electron-phonon coupling constant of Eq.~\eqref{eq:coupling} now takes the form
\begin{equation}
    \lambda^{\rm{new}} =  \sum_{\KK,\qq,\nu,\nu'}\frac{2}{\omega_{\qq}N(0)}  |g_{\KK \qq}^{\nu \nu'}|^2
                   \delta(\xi_{\KK}^{\nu})\delta(\xi_{\KK + \qq}^{\nu'}).
    \label{eq:foldedcoupling}
\end{equation} 
The sum runs over all momenta $\KK$ inside the reduced Brillouin zone, and all $\qq$ vectors in the Brillouin zone of the pristine material. This is equivalent to allowing umklapp scattering, but only within the first Brillouin zone of the pristine material, and allows to make a meaningful comparison with the pristine material.

Qualitatively, to achieve the highest coupling constant $\lambda$, we need to move weight from the original Fermi surface to match with the points of strongest electron-phonon coupling. For example, in many materials the interaction matrix element favors certain large scattering vectors~\cite{Pickett2008}. It is then beneficial to allow many processes, where these wave vectors can scatter. In our example, we achieve precisely that: because the shadow bands cover much more area in the Brillouin zone, more phase space is available where it matters. This is relevant for materials like MgB$_2$, where the coupling is strong for a small region in $\qq$ space only, and giving more space in $\kk$ space would be beneficial. In such a situation, the nanopatterning  also weakens the Kohn anomaly and thus increases the phonon energy, further benefitting the superconductivity~\cite{Pickett2006}.

\begin{figure}[htb] 
\centering
      \includegraphics[width=1 \textwidth]{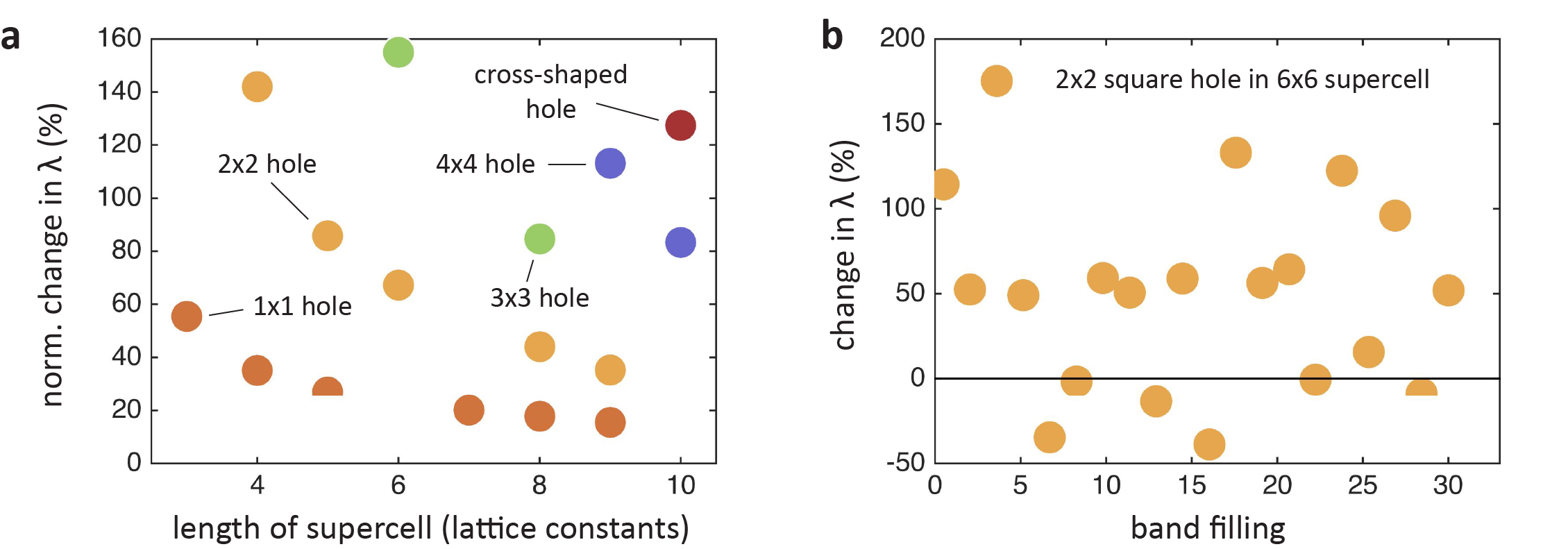}
      \caption{Enhancement of $\lambda$ and $T_c$. \figla{a}, Dependence of the coupling constant $\lambda$  on the supercell size $L$ and shape, normalized by the change in $N(0)$ (see Appendix for details). The labels are for all markers with the same color. All values are relative to the pristine material and normalized for density of states effects. The filling in both pristine and meta-superconductor is 0.5\% for all supercell sizes, to ensure that we are below the first band gap and thus concentrate on the kinematic constraints.   \figla{b} Dependence of the coupling constant $\lambda$ on the filling. Here, the resulting change is a combination of density of states effects and kinematic effects. }\label{fig:con} 
\end{figure}

Figure ~\ref{fig:con}\figla{a} summarizes our results for different supercell shapes and sizes. Even for larger supercell sizes, coinciding with what is presently possible with nanofabrication, one can improve the electron-phonon coupling by `aligning' the kinematic constraints. Further, there is a clear dependence on the shape of the supercell.
Finally, a strong effect stems from the increased  density of states at certain energies due to the opening of band gaps. Fig.~\ref{fig:con}\figla{b} shows the total coupling as a function of filling. The enhancements/suppressions at different fillings stem from density of states effects and from alterations of the phase space.

The electronic structure is not the only driver of the coupling. The periodic supercell structure can have a strong effect on the phonons as well. Designing phonons is well studied in opto-mechanics~\cite{Safavi2014} and sonic engineering~\cite{Deymier2013}, but in the Eliashberg approximation they do not significantly influence the overall coupling~\cite{Marsiglio2008}. More accurate calculations that take the retardation effects better into account go beyond the scope of this paper, but we note that recent Monte Carlo simulations revealed a strong dependence of the phonon energy on the superconducting transition temperature~\cite{Lin2016}, showing the additional potential of phonon engineering. We further mention that important effects such as the suppression of competing orders (e.g., charge density waves), or the improvement of screening~\cite{Shen2005} similar to recently proposed colloid arrays~\cite{Smolyaninov2013, Smolyaninov2016} could yield further opportunities for improved superconductors. 

While a model calculation as presented here allows to make educated guesses for a nano-patterning strategy on real materials, more sophisticated approaches can be used for specific materials. Local density approximation calculations can determine the exact phononic and electronic structure of simple materials, with supercell sizes in the range of several tens of lattice sites, and even take into account effects from dangling bonds. Second, finite-element calculations in combination with a tight-binding approach have proven to be a powerful tool to calculate the phononic structures at very low frequencies~\cite{Deymier2013, Safavi2014}. For materials, where these branches are known to dominate electron-phonon coupling, this approach, generalized also for the electronic structure, will be useful. Third, new theoretical models that directly connect  available parameters like ultrasonic attenuation with the coupling parameter $\lambda$ might be useful for materials, where less is known about the underlying microscopic structure. Importantly, these methods should allow searches in a wide parameter space for specific materials.  

To conclude this paper we would like to note that `natural' nanoscale patterning has been seen in many quantum materials~\cite{Gunnarsson1996,Fujita2012,Comin2016,Maeno1998,Cohen1968,Geim1997}.  (The effects described in this Letter are weaker but present in structures with clear spatial correlation length.) It is known that, depending on the lenghtscale, granular aluminum~\cite{Cohen1968,Geim1997}  
can have enhanced $T_c$ compared to the constituents, but the cause for this is still debated.  In high-temperature superconductors, a granular structure seems to be native to the material~\cite{Fujita2012,Comin2016}. Further, some interface superconductors can show superstructures due to Moir\'e effects. Finally, in fullerene superconductors, the C$_{60}$ molecules form native supercells~\cite{Gunnarsson1996}. Some of these superconductors might not be driven by  phonons, but most of our calculations are rather independent of the character of the mediating bosons. Whether there is a connection between these materials and the method presented here, and whether artificially designed granularity can be superior to the current state in these materials remains to be seen.


\section*{Acknowledgements}
We are thankful to Vadim Cheianov for guidance and to Jan Aarts,  Irene Battisti, Koen Bastiaans,  Andrea Caviglia, J.C.\ S\'eamus Davis, John P.\ Davis, Simon Gr\"oblacher, Alex Kemper,     Amir Safavi-Naeni, Manfred Sigrist, Ronny Thomale, Jasper van Wezel for helpful discussions.
\paragraph{Funding information}
We acknowledge support from the Netherlands Organization for Scientific Research (NWO) as part of the Frontiers of Nanoscience programme and the Vidi talent scheme (M.P.A.) and   the Swiss Society of Friends of the Weizmann Institute (M.H.F).
\begin{appendix}

\appendix

\section{Appendix: Model calculation}
In this  {Appendix}, we describe in more detail our model calculation for a square lattice with a nano-patterned supercell.  We use the following  strategy: 

\begin{enumerate}
\item 
For illustration and for comparison, we first consider the model for the pristine material with the coupling matrix element $g^{0}_{\kk\qq}$, the resulting coupling constant $\lambda$, and the transition temperature $T_c$.
\item 
We model the supercell by setting the electron hopping to zero to the sites that are part of the hole, and find the new basis in which the electron Hamiltonian is diagonal. Further, we find the eigenvalues and transformation matrices.
\item 
We take the pristine interaction Hamiltonian, but replace the original electron  operators by the new basis. This allows us to obtain the new interaction matrix element $g^{\rm{\nu\nu'}}_{\KK\qq}$ between the new eigenstates. Here, we ignore the phonons because the details of their dispersion is not crucial in BSC theory. 
\item
We calculate the coupling constants $\lambda$ from the new interaction matrix elements and electron dispersions.
\end{enumerate}

\subsection{Model for the pristine material} \label{pris}

We start by describing the model for the pristine material (before the nano-patterning),  a two-dimensional square lattice with $N\times N$ sites  and Hamiltonian 
\begin{equation}
\HH=\HH_{\rm{el}}+\HH_{\rm{ph}}+\HH_{\rm{el-ph}},
\end{equation} 
where $\HH_{\rm{el}}$, $\HH_{\rm{ph}}$, $\HH_{\rm{el-ph}}$ are the electronic, phononic, and interaction parts of the Hamiltonian, respectively.

For the electronic part, we use a tight-binding description on the ions' equilibrium positions assuming nearest-neighbour hopping only with hopping element $t$,  
\begin{equation}
    \HH_{\rm{el}}=-t\sum_{(\rr, \rr')} c_\rr^\dagger c^{\phantom{\dag}}_{\rr'} - \mu \sum_{\rr}c_\rr^\dagger c^{\phantom{\dag}}_\rr,
    \label{eq:Hel}
\end{equation} 
where $c_\rr^\dagger$  ($c^{\phantom{\dag}}_\rr$) is the electron creation (annihilation) operator, $\mu$ is the chemical potential and $\rr$ denote the lattice sites. The electron creation operators in momentum space are 
\begin{equation}
    c_\kk ^\dagger = \frac{1}{N} \sum_\rr e^{-i\kk\cdot\rr}c^{\dag}_{\rr},
\end{equation}
with
\begin{equation}
    k_{x},  k_{y}\in \{ -\frac{\pi}{a}, -\frac{\pi}{a}\frac{N-1}{N}, \ldots \frac{\pi}{a}\frac{N-1}{N}  \},
\end{equation}
where $a$ is the lattice constant (in the following, we set $a=1$). These operators diagonalize the electronic Hamiltonian Eq.~\eqref{eq:Hel},
\begin{equation}
    \HH_{\rm{el}}= \sum_{\kk} \varepsilon^\pdag_{\kk}c^\dag_{\kk}c^{\phantom{\dag}}_{\kk},
\end{equation} 
with $\varepsilon_{\kk}=-2t(\cos k_x + \cos k_y)-\mu$ the  electron dispersion.

We consider acoustic phonons stemming from nearest-neighbor ($\kappa$) and next-nearest-neighbor ($\kappa'$) springs on a square lattice,   
\begin{equation}
    \HH_{\rm{ph}}=\sum_{\rr}\frac{\v{p}_\rr^2}{2m}  + \frac{\kappa}{2} \sum_{(\rr,\rr')} \big(\v{e}_{nn}\cdot(\uu_\rr - \uu_{\rr'}))\big)^2
    +  \frac{\kappa'}{2} \sum_{[\rr,\rr']} \big(\v{e}_{nnn}\cdot(\uu_\rr - \uu_{\rr'})\big)^2  +  \frac{\kappa''}{2} \sum_{\rr}  \uu_\rr ^2,
\end{equation} 
where $\v{p}_\rr$ are the ion momenta, $\uu_\rr$ are the deviations from equilibrium of the ions at site $\rr$, $m$ is the ion mass, and $\kappa$, $\kappa'$ are the spring constants.  $\v{e}_{nn}$ and $\v{e}_{nnn}$ denote the unit vectors in the direction of nearest and next-nearest neighbours, respectively.  To facilitate numerics, we added a ``mass-term'' $\kappa''$ that removes the modes with zero energy. In the above sum, $(\rr,  \rr')$  denotes nearest neighbours and $[\rr,  \rr']$  denotes next-nearest neighbours. 
It is convenient to write the potential part of the phonon Hamiltonian in matrix form, 
\begin{equation}
\HH_{\rm{ph}}= \sum_\qq
 \begin{pmatrix*}
 u_\qq^x & u_\qq^y
 \end{pmatrix*}
 \mat{H}
 \begin{pmatrix*}
 u_{-\qq}^x \\ u_{-\qq}^y
 \end{pmatrix*},
\end{equation}
with the matrix  
\begin{equation}
  \mat{H}=
 \begin{bmatrix*}
2\kappa \sin^2 (\frac{q_x}{2}) +\kappa' (1-\cos {q_x}\cos{q_y} ) + \kappa''& \kappa'\sin q_x \sin q_y 
\\  \kappa'\sin q_x \sin q_y & 2\kappa \sin^2 (\frac{q_y}{2}) +\kappa' (1-\cos {q_x}\cos{q_y} ) +\kappa''
 \end{bmatrix*}.
\end{equation}
Diagonlaizing this matrix leads to the phonon spectrum
\begin{equation}
\begin{aligned}
 \omega^{\pm}_\qq &= &\frac{2}{m} \Big[ \kappa  \sin^2 \big(\frac{q_x}{2}\big)+\kappa   \sin^2 \big(\frac{q_y}{2}\big)
  + \kappa' \big( 1-\cos q_x \cos q_y \big)+\kappa'' \Big] \\ &&\pm \frac{2}{m}  \sqrt{ \frac{\kappa^2}{4} (\cos q_y -\cos q_x)^2   + \kappa'^2 \sin^2 (q_x)  \sin^2(q_y)}.
  \end{aligned}
 \end{equation}
The (normalized) eigenvectors of the matrix, $\v{e}_{\qq}^\pm$,  indicate the polarisations of the normal modes. Along the high-symmetry axes (the nearest and next-nearest neighbour directions), as well as in the limit $q=|\qq| \ll 1$, $\v{e}_{\qq}^+$ points exactly along the \emph{longitudinal} direction, and $\v{e}_{\qq}^-$ along the \emph{transversal} direction. For the parameters considered here, the deviation of  $\v{e}_{\qq}^+$ from the longitudinal direction is small:  $\v{e}_{\qq}^+\cdot\qq/q\geq 0.95$ throughout the  Brillouin zone. We thus approximate $\v{e}_{\qq}^+ \approx \v{e}_{\qq}^{\rm{long}}=\qq/q$.
 
The phonon Hamiltonian can be written in second-quantized form by introducing the phonon creation and annihilation operators $a^{\phantom{\dag}}_{\qq,\pm}$, $a^\dagger_{\qq,\pm}$. The displacement of the ion at site $\rr$ is then
\begin{equation}
\begin{aligned}
    \uu_\rr &= \frac{1}{N}\sum_{\qq,\pm}\v{e}_\qq^{\pm} e^{i\qq\cdot\rr} u^{\pm}_\qq \\ & = \frac1N\sum_{\qq,\pm} \v{e}_\qq^{\pm} e^{i\qq\cdot\rr}\Big(\frac{\hbar}{2m\omega^{\pm}_\qq}\Big)^{\frac{1}{2}} \big(a^{\phantom{\dag}}_{\qq, \pm} + a^\dagger_{-\qq, \pm}\big).
    \label{eq:Ur}
    \end{aligned}
\end{equation} 
 
We introduce the coupling of the electrons to the lattice by considering the energy change of the electronic states when the background (ion) density changes as the crystal expands/contracts, analogous to the dependence of the chemical potential on the density for a free electron gas. This leads to the interaction Hamiltonian 
\begin{equation} \label{eq:int}
    \HH_{\rm{el-ph}}  =   {D}  \sum_\rr   \frac{\Delta V_\rr}{V}  c^\dagger_\rr c^{\phantom{\dag}}_\rr.
\end{equation} 
The constant ${D}$ indicates the proportionality between change of the chemical potential and volume change  $\Delta V /V$; it is commonly called \emph{displacement potential}.  For acoustic phonons, the volume change is approximately given by the divergence of the displacement field, 
\begin{equation} 
\frac{\Delta V_\rr}{V} \approx {\nabla}\cdot \uu_\rr.
\end{equation} 
Using the expression in Eq.~\eqref{eq:Ur}, we can write
\begin{equation} \label{eq:divUr}
\begin{aligned}
    \nabla \cdot \uu_\rr & =  \frac1N\sum_{\qq,\pm}\nabla\cdot \v{e}_\qq^{\pm} e^{i\qq\cdot\rr}\Big(\frac{\hbar}{2m\omega^{\pm}_\qq}\Big)^{\frac{1}{2}} \big(a^{\phantom{\dag}}_{\qq, \pm} + a^\dagger_{-\qq, \pm}\big)
   \\ 
   & =    \frac iN\sum_{\qq,\pm} \qq\cdot\v{e}_\qq^{\pm} e^{i\qq\cdot\rr}\Big(\frac{\hbar}{2m\omega^{\pm}_\qq}\Big)^{\frac{1}{2}} \big(a^{\phantom{\dag}}_{\qq, \pm} + a^\dagger_{-\qq, \pm}\big)
   \\ 
   & \approx    \frac iN\sum_{\qq}q e^{i\qq\cdot\rr}\Big(\frac{\hbar}{2m\omega^{+}_\qq}\Big)^{\frac{1}{2}} \big(a^{\phantom{\dag}}_{\qq,+} + a^\dagger_{-\qq,+}\big),
\end{aligned}
\end{equation} 
where we approximated  $\qq\cdot\v{e}_\qq^{+} \approx q$ and $\qq \cdot \v{e}_\qq^{-} \approx 0$ as described above, i.e.\ only the phonons with energy $\omega_\qq^+$, which are approximately the longitudinal phonons, couple to electrons, in accordance to Adler's theorem. For simplicity of notation, we will drop the `$+$' in the following. Note that there are different models that lead to  a similar coupling Hamiltonian; overviews can be found in Refs~\cite{Madelung1978,  Bruus2004, Phillips2012, Coleman2015}. We also note that in the case of the lattice considered there, the divergence is of the form $\nabla \cdot \uu_\rr  \approx [\sin^2q_x u_q^x + \sin^2q_y u_q^y]$, however, as this will not change the ratios displayed in Fig.~4 of the main text, we continue with the continuum approximation above.
Inserting Eq.~\eqref{eq:divUr} into Eq.~\eqref{eq:int} and using $\qq=\kk-\kk'$ yields 
\begin{equation}
\begin{aligned}
    \HH_{\rm{el-ph}}& =    {D}  \sum_\rr   \frac{\Delta V_\rr}{V}  c^\dagger_\rr c^{\phantom{\dag}}_\rr  \\
    & = \frac {i}{N^3}\sum_\rr \sum_{\kk\qq} e^{i (\qq+\kk-\kk')\cdot\rr} D_\qq \Big(\frac{\hbar}{2m\omega_\qq}\Big)^{\frac{1}{2}} \big(  a_\qq + a_{-\qq}^\dagger \big)c_{\kk'}^\dagger c^{\phantom{\dag}}_{\kk} \\
    &=  \frac {i}{N}\sum_{\kk\qq} g_{\kk\qq}^0 \big(a_\qq + a^\dagger_{-\qq}\big)c_{\kk+\qq}^\dagger c^{\phantom{\dag}}_\kk ,
    \label{eq:elphcoupl}
\end{aligned}
\end{equation} 
with the coupling matrix element
\begin{equation}
    g_{\kk\qq}^0=g_{\qq}^0={D}q\Big(\frac{\hbar}{2m\omega_\qq}\Big)^{\frac{1}{2}} = D_\qq\Big(\frac{\hbar}{2m\omega_\qq}\Big)^{\frac{1}{2}},
\end{equation} 
where we also introduced the momentum dependent proportionality  $D_\qq$. In the simple approximation above, it is   given by  the expression above with  ${D}$ being a  constant, but more generally it is related to the Fourier transform of the atomic potential, leading to different, material-specific expressions.

Given the interaction matrix element in Eq.~\eqref{eq:elphcoupl}, the dimensionless coupling parameter $\lambda$ can be expressed as
\begin{equation}
    \lambda = \frac{1}{N^4} \sum_{\kk\qq}\frac{2}{\omega_{\qq}N(0)}  |g_{\qq}^0|^2
                   \delta(\varepsilon_{\kk})\delta(\varepsilon_{\kk + \qq}).
\end{equation} 
The product of delta functions $\delta(\varepsilon_{\kk})\delta(\varepsilon_{\kk + \qq})$ ensures that only electrons at the Fermi level contribute (we assume that phonon energies are much smaller than the Fermi energy), yielding a  kinematic constraint. It is often instructive to calculate the $\qq$ or $\kk$ dependence of the coupling constant, $\lambda_{\qq}$ and $\lambda_{\kk}$, by  summing over all other variables. This yields the contributions of a given phonon mode $\qq$  or  the contributions of a given electronic state  $\kk$ to the coupling constant $\lambda$.  

Finally, we can calculate the transition temperature $T_c$ using the standard Bardeen-Cooper-Schrieffer (BCS) theory or using the Allen-Dynes approximation in Eliashberg theory. In BCS, we have the standard exponential dependence, 
\begin{equation}\label{eg:tc1bcs}
T_c=1.13 \omega_D e^{-\frac{1}{\lambda}},
\end{equation} 
 where $\omega_D$ is a measure of the phonon energy, usually taken to be the Debey energy. In Eliashberg theory we can approximate the transition temperature by 
\begin{equation}\label{eg:tc1eliashberg}
k_B T_c=\frac{\hbar \omega_{\rm{ln}}}{1.2}\exp\Big( \frac{1.04 (1+\lambda)}{\lambda - \mu^*(1+0.62\lambda)} \Big),
\end{equation} 
where $\mu^*$ is the Coulomb pseudopotential.  $\omega_{\rm{ln}}$ is referred to as `logarithmic average' of the phonon energy, defined by 
\begin{equation}\label{eq:tc1cpseudopot}
\omega_{\rm{ln}} = \exp\Big( \frac{2}{\lambda} \int_0^{\infty} d\nu \ln(\nu) \frac{\alpha^2 F(\nu)}{\nu}\Big),
\end{equation} 
with
\begin{equation}\label{eq:tc2}
    \alpha^2 F(\nu) = \frac{1}{N^4}\frac{1}{N(0)} \sum_{\kk\qq} \delta(\nu-\omega_\qq)
 |g_{\kk\qq}^0|^2 \delta(\varepsilon_{\kk})\delta(\varepsilon_{\kk + \qq}) . 
\end{equation} 

\subsection{Electron part of the Hamiltonian with a supercell}

We now consider a supercell with $L\times L$ sites as discussed in the main text. First, we introduce some nomenclature: We denote the number of supercells with  $M^2=(N/L)^2$, each containing   $L^2$ ions, giving the same total number of atomic sites as our pristine model, $N^2$. We indicate the equilibrium position of  each ion  with $\rr=\RR + \ttau$, where $\RR$ is the position of the  supercell and  $\ttau$ is the position within the supercell.
As before, the deviation of the ions from their equilibrium position is denoted by $\uu_\rr$ and the electron creation (annihilation) operators by $ c_\rr ^\dagger$ ($ c^{\phantom{\dag}}_\rr$). 
To introduce the supercell, we now allow for arbitrary chemical potentials $\mu_{\ttau}$ at site $\ttau$ inside the supercell and arbitrary hopping constants $t_{\ttau,\ttau'}$ for neighbouring sites $\ttau$ and $\ttau'$ within the supercell, all  preserving the $L$-periodicity, 
\begin{equation}
    \HH_{\rm{el}}=-\sum_{(\rr, \rr')} t_{\ttau, \ttau'}^\pdag c_\rr^\dagger c^{\phantom{\dag}}_{\rr'} - \sum_{\rr}\mu_{\ttau}^\pdag c_\rr^\dagger c^{\phantom{\dag}}_\rr.
\end{equation} 

We can bring this Hamiltonian in block-diagonal form by introducing a Fourier transform of the electron operators with respect to the new periodicity, 
\begin{equation} \label{eq:trans}
    c_{\{\rr=\RR+\ttau\}} = \frac{1}{M} \sum_\KK e^{i\KK\cdot\RR} c_\KK^{\ttau}.
\end{equation}
Note that here and in the following, we use capital letters $\KK = (K_x,K_y)$ to denote the reciprocal wave vectors with respect to the supercell periodicity $L$, i.e., 
$$K_x , K_y\in \{ -\frac{\pi}{La}, -\frac{\pi}{La}\frac{M-1}{M},  \ldots \frac{\pi}{La}\frac{M-1}{M}  \}.$$  
This leads to the block diagonal Hamiltonian
\begin{equation}
\HH_{\rm{el}}=\sum_\KK \ov{c}_\KK^\dagger [\mat{H}_\KK]\ov{c}_\KK^{\phantom{\dag}},
\end{equation} 
where we introduced the vectors of operators $\ov{c}_\KK = (c_\KK^1, c_\KK^2, \ldots c_\KK^{L\times L})$.   For simplicity, when used as an index, we write $\tau$ instead of a  vector $\ttau$ for the position within the supercell. 

Each  block Hamiltonian $[\mat{H}_K]$ is an $L^2\times L^2$ matrix.  It contains diagonal elements with the chemical potentials of all $L^2$ sites, all the hopping elements, as well as the phase factors for connections between sites in adjacent supercells. Specifically, its elements are 
\begin{equation}\label{blockH}
\begin{aligned}  
 [\mat{H}_\KK]^{\tau \tau'} &=- \mu_\tau \delta_{\tau \tau'}
	 -t_{\tau, \tau'}  \delta_{(\tau, \tau')}  -t_{\tau ,\tau'}  \delta_{(\tau, \tau')}  \\
         & - t_{\tau,\rm{up}}\delta_{(\tau,\rm{up})}e^{-iLK_y} - t_{\tau,\rm{down}}\delta_{(\tau,\rm{down})}e^{iLK_y} \\
                  & - t_{\tau,\rm{right}}\delta_{(\tau,\rm{right})}e^{-iLK_x} - t_{\tau,\rm{left}}\delta_{(\tau,\rm{left})}e^{iLK_x} ,
\end{aligned}  
\end{equation} 
where again we use a single index to denote the position in the supercell, and $ \delta_{\tau \tau'}=1$ only when $ \tau$ and  $\tau'$ are the same while $ \delta_{(\tau \tau')}=1$ when $ \tau$ and  $\tau'$ are nearest neighbours. $\delta_{(\tau,\rm{right})}=1$ when $\tau$ is nearest neighbour to a site in the next supercell to the right, and similar for left, up and down.
It is instructive to look at this in  one dimension, in which case the  block Hamiltonian $[\mat{H}_\KK]$ is an $L\times L$ matrix:
\begin{equation}
 [\mat{H}_\KK] = 
 \begin{pmatrix*}
 - \mu_1 &-t_{12} &  &   &-t_{1,L} e^{-i(La)K} \\
 -t_{21}  & -\mu_2  &  &    &\\
  &   & \ddots &  &\\
   &   &  & -\mu_{L-1}& -t_{L-1,L}  \\
   -t_{L,1} e^{i(La)K} &  &   & -t_{L,L-1} &-\mu_L 
\end{pmatrix*}.
\end{equation}

Finally, diagonalizing the matrices $[\mat{H}_\KK]$ for each $\KK$ yields the eigenstates with operators 
\begin{equation}
    {\gamma}^{\nu}_\KK = \sum_\tau[\mat{U}_{\KK}]^{\nu\tau} c_\KK^{\tau},
\end{equation} 
with transformation matrices $[\mat{U}_\KK]^{ \nu \tau}$ and the corresponding eigenenergies $\xi_\KK^\nu$ such that 
\begin{equation}
    \HH_{\rm{el}} = \sum_\KK\sum_\nu \xi^\nu_\KK (\gamma^{\nu}_\KK)^{\dagger}  \gamma^\nu_\KK.
\end{equation} 

It is possible to implement any kind of supercell with different chemical potentials $\mu_\tau$ and hopping $t_{\tau \tau'}$ in our calculation. To model the specific hole that we describe in the main text, we first designate some sites as part of the hole. These  sites have zero hopping probability to all neighbours, leaving the states completely non-dispersive. Then, we increase the chemical potential at all hole sites to move the non-dispersive states above the relevant bands and thus ensure that only the latter contribute to superconductivity.

\subsection{Phonon part of the Hamiltonian with a supercell}
Structures with periodic patterning made to alter phonon dispersions have been created in the context of opto-mechanics, where they are known as `phononic crystals'~\cite{Safavi2014}, or in the context of sound insulation and engineering, where  they are known as  `sonic crystals' or `acoustic metamaterials''~\cite{Deymier2013}.)  

While the periodically patterned films we propose  influence both phononic and electronic structure, the fabrication method allows us to  chose which one we influence strongest. In this Letter, we concentrate on the electronic structure, as in general  the critical temperature within the BCS framework is not influenced by the momentum dependence of the phonon energies; models that correctly take phonon folding into account go  beyond the scope of this paper. 

The calculations of the folding of phonons  using distributed mass-spring elements is  similar to what we describe for electrons, and it has been done in the context described above. For detailed description of the formalism see e.g.\  Refs~\cite{Jensen2003, Cao2009, Deymier2013}. (One can also calculate the phonon dispersions in the elastic approximation~\cite{Safavi2014}.)
For the model in the main text we require a hole in the pristine material.  The spring constants to the sites that are part of the hole are set to be small compared to the spring constant of the pristine material, to avoid coupling, and the masses to be heavy to suppress movements.

\subsection{The electron-phonon coupling in a system with supercells}

We now want to write the new interaction matrix element $g^{\rm{\nu\nu'}}_{\KK\qq}$ of our meta-material as a function of the folded electronic structure and interaction matrix element $g^0_{\qq}$ of the pristine material.

Our starting point is again the interaction Hamiltonian of the form 
\begin{equation} 
    \HH_{\rm{el-ph}} =    \sum_{\rr} \underbrace{ D\vec{\nabla}\cdot \v{u_\rr^\pdag}}_{\rm (I)} \underbrace{ c_\rr^\dagger c_\rr^\pdag }_{\rm (II)}.
\end{equation} 
As in the pristine case, we write for the phonon part 
\begin{equation} \label{eq:div}
\begin{aligned}
    \rm{(I)} & = D \nabla \cdot \uu_\rr \\
    &\approx \frac iN\sum_\qq g_\qq^0  e^{i\qq\cdot\rr}  \big(a^{\phantom{\dag}}_\qq + a^\dagger_{-\qq}\big).
\end{aligned}
\end{equation} 
However, we now replace the electron operators of the pristine material with the operators of the new eigenstates.
The real space electron operators are then given by
\begin{equation}
    c_{\{\rr=\RR+\ttau\}}  = \frac{1}{M} \sum_\KK \sum_\nu e^{i\KK\cdot\RR} [\mat{U}^\dagger_\KK]^{\tau \nu} \gamma_\KK^\nu.
\end{equation} 

This yields an electron part
\begin{equation}
\begin{aligned}
\rm{(II)} &=    c_\rr^\dagger c_\rr   \\
& =    \frac{1}{M^2} \sum_{\KK\KK'} \sum_{\nu\nu'} \Big[ e^{i\KK\cdot\RR} (\gamma_\KK^{\nu})^{ \dagger} \mat{U}_\KK^{\nu\tau} \Big]  \Big[ e^{-i\KK'\cdot\RR} [\mat{U}_{\KK'}^\dagger ]^{\tau\nu'} \gamma_{\KK'}^{\nu'}  \Big]
  \\
  & =   \frac{1}{M^2} \sum_{\KK\KK'} \sum_{\nu\nu'} e^{i(\KK-\KK')\cdot\RR}[\mat{U}_\KK]^{\nu\tau}    [ \mat{U}^\dagger_{\KK'} ]^{\tau\nu'}(\gamma_\KK^{\nu})^{\dagger} \gamma_{\KK'}^{\nu'}.
\end{aligned}
\end{equation}  
Putting both together and using 
\begin{equation}
\frac{1}{M^2}\sum_\RR e^{i(\qq-\KK+\KK')\cdot\RR} = \sum_{\bl} \delta_{\KK,\KK'+\QQ+\frac{2\pi}{La}\bl}, 
\end{equation}
(introducing $\qq = \QQ + \frac{2\pi}{La}\bl$)
we obtain  
\begin{equation}
\begin{aligned}
    \HH_{\rm{el-ph}}& = \sum_\rr D_\qq (\vec{\nabla}\cdot \v{u}_{\rr}) c^\dagger_\rr c^{\phantom{\dag}}_\rr   \\
    & =  \frac{i}{M^2 N}\sum_\rr  \sum_{\KK\KK'}\sum_{\QQ,\bl} \sum_{\nu\nu'} e^{i(\KK-\KK'-\QQ)\cdot\RR} [\mat{U}_{\KK}]^{\nu\tau}    [\mat{U}_{\KK'}^\dagger ]^{\tau\nu'}(\gamma_\KK^\nu)^\dagger \gamma_{\KK'}^{\nu'}  \\
  & \times  g_\qq^0 e^{i\qq \cdot \ttau} \big(a_\qq^\pdag + a^\dagger_{-\qq}\big) \\
  & = \frac iN \sum_{\KK\QQ}\sum_{\bl} \sum_{\ttau}  \sum_{\nu\nu'}g_\qq^0 e^{i\qq \cdot \ttau} 
  [\mat{U}_{\KK+\QQ}]^{\nu\tau}    [\mat{U}_{\KK}^\dagger ]^{\tau\nu'}(\gamma_{\KK+\QQ}^\nu)^\dagger \gamma_{\KK}^{\nu'}    \big(a_\qq + a^\dagger_{-\qq}\big).
\end{aligned}
\label{eq:intelfolded}
\end{equation}  
Note that the sum runs over all $\KK$ inside the reduced Brillouin zone, and all $\qq$ vectors in the Brillouin zone of the pristine material. This is equivalent to allowing umklapp scattering, but only within the Brillouin zone of the pristine material, to make a meaningful comparison with the pristine material. 

Finally, we rewrite the interaction Hamiltonian as
\begin{equation}
    \HH_{\rm{el-ph}} =   \frac iN  \sum_{\KK \QQ, \bl}   \sum_{\nu \nu'}  g^{\nu\nu'}_{\KK\qq} \,\, (\gamma_{\KK+\QQ}^\nu)^\dagger \gamma_{\KK}^{\nu'}    \big(a_\qq^\pdag + a^\dagger_{-\qq}\big),
\end{equation}  
with the coupling vertex for the new eigenstates
\begin{equation}\label{eq:vertexSC}
g^{\nu\nu'}_{\KK\qq}  =  g^0_\qq    \sum_{\tau}  e^{i\qq \cdot \ttau} 
  [\mat{U}_{\KK+\QQ}]^{\nu\tau}    [\mat{U}_{\KK}^\dagger ]^{\tau\nu'}.
\end{equation}  
Again,  the displacement potential proportionality can have a complex, material-specific $\qq$ dependence, i.e.\ ${D}$ does not need to be constant. This stems from the spatial dependence of the deformation potential, and reflects the interaction matrix element in real materials such as MgB$_2$.

\subsubsection{Coupling parameter $\lambda$ and transition temperature $T_c$}
We calculate the coupling parameter $\lambda$ similar to the case of the pristine material,  where we used
\begin{equation}
    \lambda^{\rm{pristine}} =  \frac{1}{N^4}\sum_{\kk\qq}\frac{2}{\omega_{\qq}N(0)}  |g_{\qq}^0|^2
                   \delta(\varepsilon_{\kk})\delta(\varepsilon_{\kk + \qq}),
\end{equation} 
but now replacing the interaction matrix element from the pristine material, $g_{\qq}^0$, with the one calculated above, $g_{\KK\qq}^{\nu \nu'}$, and summing over momenta $\qq$ in the `large' Brioullin zone of the pristine material. As mentioned, one can interpret this as including an amplitude for umklapp scattering in between the new, `small' Brioullin zones. This yields
\begin{equation}\label{eq:lambdaSC}
    \lambda^{\rm{SC}} =  \frac{1}{N^4}\sum_{\KK\qq\nu\nu'}\frac{2}{\omega_{\qq}N(0)}  |g_{\KK \qq}^{\nu \nu'}|^2
                   \delta(\xi_{\KK}^{\nu})\delta(\xi_{\KK + \qq}^{\nu'}),
\end{equation} 
with $ g_{\KK\qq}^{\nu \nu'}$ defined above. Note that this is now a multi-band superconductor. Throughout this work, we use the coupling parameter $\lambda$ as a figure of merit. We emphazise that it is directly related to the  transition temperature $T_c$  according to Eqs~\ref{eg:tc1bcs}-\ref{eq:tc2}.

\subsubsection{Numerical implementation}
We use the commercial Matlab package for all calculations, and we will now briefly outline how all calculations  can be expressed as combinations of matrix operation and matrix diagonalization.  
 We start with the interaction matrix element from Eq.~\eqref{eq:vertexSC} and insert it into Eq.~\eqref{eq:lambdaSC}:
\begin{equation}
   \begin{aligned}
 \lambda^{\rm{SC}}& =   \frac{1}{N^4}\sum_{\KK\qq\nu\nu'}\frac{2}{\omega_{\qq}N(0)}  |g_{\KK\qq}^{\nu \nu'}|^2
                   \delta(\xi_{\KK}^{\nu})\delta(\xi_{\KK + \qq}^{\nu'}) 
                   \\
                 &  =  \frac{1}{N^4}\sum_{\KK\qq\nu\nu'}\frac{2}{\omega_{\qq}N(0)}  
                 \Bigg| g_\qq^0    \sum_{\tau}  e^{i\qq \cdot \ttau} 
  [\mat{U}_{\KK+\qq}]^{\nu\tau}    [\mat{U}_{\KK}^\dagger ]^{\tau\nu'}   \Bigg|^2
   \delta(\xi_{\KK}^{\nu})\delta(\xi_{\KK + \qq}^{\nu'})
  \\
  &=   \frac{1}{N^4} \sum_{\KK\qq}\frac{2}{\omega_{\qq}N(0)}  |g_\qq^0|^2  
  \sum_{\nu\nu'} \Bigg| \sum_{\tau}  e^{i\qq \cdot \ttau} 
  [\mat{U}_{\KK+\qq}]^{\nu\tau}    [\mat{U}_{\KK}^\dagger ]^{\tau\nu'} \delta(\xi_{\KK}^{\nu}) \Bigg|^2 \delta(\xi_{\KK + \qq}^{\nu'})
  \\
  &=   \frac{1}{N^4} \sum_{\KK\qq} F_\qq   
  \sum_{\nu\nu'} \Big| \sum_{\tau}  e^{i\qq \cdot \ttau} 
  [\mat{U}_{\KK+\qq}]^{\nu\tau}    [\mat{U}_{\KK}^\dagger ]^{\tau\nu'} \Big|^2 \delta(\xi_{\KK}^{\nu})  \delta(\xi_{\KK + \qq}^{\nu'}),
 \end{aligned}
\end{equation}  
with 
\begin{equation}
F_\qq = \frac{2}{\omega_{\qq}N(0)}      \frac{D_\qq^2 \hbar }{2m\omega_\qq} = |g_\qq^0|^2 \frac{2}{\omega_{\qq}N(0)} .
\end{equation}

In the last two steps we used the fact that the terms we took outside the sum are positive and independent of $\nu\nu'$. 
We further define $\lambda_{\KK\qq}$ such that $\lambda^{\rm{SC}} =  1/(N^2M^2) \sum_{\KK\qq}\lambda_{\KK\qq}$. 

Now we can rewrite this into a form convenient for numerical matrix operations:
\begin{equation}
\begin{aligned}
    \lambda_{\KK\qq} &=   \frac{F_\qq}{L^2}   
  \sum_{\nu\nu'} \Big| \sum_{\tau}  e^{i\qq \cdot \ttau} 
  [\mat{U}_{\KK+\qq}]^{\nu\tau}    [\mat{U}_{\KK}^\dagger ]^{\tau\nu'} \Big|^2 \delta(\xi_{\KK}^{\nu})  \delta(\xi_{\KK + \qq}^{\nu'})
\\
  & =\frac{F_\qq}{L^2} 
  \sum_{\nu\nu'} \Big| \sum_{\tau\tau'}  
   [\mat{U}_{\KK+\qq}]^{\nu\tau}   [\delta_{\tau\tau'} e^{i\qq \cdot \ttau}]^{\tau\tau'}[\mat{U}_{\KK}^\dag ]^{\tau'\nu'}   \Big|^2 \delta(\xi_{\KK}^{\nu})  \delta(\xi_{\KK + \qq}^{\nu'})
  \\
 & = \frac{F_\qq}{L^2}   
  \sum_{\nu\nu'} \Big|  \Big[ [\mat{U}_{\KK+\qq} ]  * [\rm{diag}(e^{i\qq \cdot \ttau})] *[\mat{U}^\dag_{\KK}] 
  \Big]^{\nu\nu'} \Big|^2 \delta(\xi_{\KK}^{\nu})  \delta(\xi_{\KK + \qq}^{\nu'})
  \\
  & = \frac{F_\qq}{L^2}   
  \sum_{\nu\nu'} \Big|  \Big[  [\mat{U}_{\KK+\qq} ]  *  [\rm{diag}(e^{i\qq \cdot \ttau})] *
  [\mat{U}_{\KK}^\dag] \Big]^{\nu\nu'}   \Big|^2 \star \Big[\delta(\xi_{\KK + \qq}^{\nu'})*\delta(\xi_{\KK}^{\nu})\Big],
  \end{aligned}
\end{equation}  
where $*$ indicates a matrix multiplication, $\star$ indicates element-wise multiplication, and  $ [\rm{diag}(e^{i\qq \cdot \ttau})]$  is a $L^2 \times L^2$ diagonal matrix with diagonal elements $(e^{i\qq \cdot \ttau})$. The sum over $\nu\nu'$ is then  the sum over  matrix elements after taking the absolute square.  The multiplication  $\delta(\xi_{\KK + \qq}^{\nu'})*\delta(\xi_{\KK}^\nu)$ is a multiplication of a $L^2\times1$ vector with a $1 \times L^2$ vector yielding a $L^2\times L^2$ matrix.  
 
For Figure 4a in the main text, we want to concentrate on kinematic effects only. We take care of a normalisation in the following way. First, we always compare to the pristine material at the same filling. Second, for the results shown in Fig.~4a, we normalize by the density of states to concentrate on the kinematic effects only.

\end{appendix}



\bibliography{refsSCAU.bib}

\nolinenumbers

\end{document}